# Street View Data Collection Design for Disaster Reconnaissance

Nicole A. Errett,[a] Joseph Wartman,[b] Scott B. Miles,[c] Ben Silver,[d] Matthew Martell,[e] Youngjun Choe[e]

[a]Department of Environmental & Occupational Health Sciences
[b]Department of Civil & Environmental Engineering
[c]Department of Human Centered Design & Engineering
[e]Department of Industrial & Systems Engineering
University of Washington, Seattle

[d]BERK Consulting, Inc

## Abstract

Over the last decade, 'street-view' type images have been used across disciplines to generate and understand various place-based metrics. However efforts to collect this data were often meant to support investigator-driven research without regard to the utility of the data for other researchers. To address this, we describe our methods for collecting and publishing longitudinal data of this type in the wake of the COVID-19 pandemic and discuss some of the challenges we encountered along the way. Our process included designing a route taking into account both broad area canvassing and community capitals transects. We also implemented procedures for uploading and publishing data from each survey. Our methods successfully generated the kind of longitudinal data that can be beneficial to a variety of research disciplines. However, there were some challenges with data collection consistency and the sheer magnitude of data produced. Overall, our approach demonstrates the feasibility of generating longitudinal 'street-view' data in the wake of a disaster event. Based on our experience, we provide recommendations for future researchers attempting to create a similar data set.

## Introduction

In January 2020, the SARS-CoV-2 virus was first detected in the United States outside of Seattle. In March 2020, the city and Washington state responded with the implementation of restrictive public health measures, including limiting the size of gatherings and issuing its first stay-at-home order. In response, a team of multidisciplinary University of Washington investigators, in partnership with the Natural Hazards Engineering Research Infrastructure (NHERI) RAPID Facility, designed and implemented a longitudinal (time-series) "street view" campaign to collect image data tracking the pandemic's impacts and the community's recovery. In May 2020, the COVID-19 Seattle Street View Campaign was launched to produce an unprecedented, high-resolution, ground-based record of the urban region during and after a major global crisis.  Since May 2020, the campaign has collected imagery data approximately every two to six weeks along a set, approximately 100-mile route through the city.

In addition to collecting data to support our team's research interests and objectives, we explicitly sought to advance the scientific application of post-event mobile imaging by establishing sampling protocols that may be used to guide campaigns for future disruptive events.  To this end, two data collection strategies were developed and adopted. The first was canvassing, an umbrella approach to acquire data that may be used to answer diverse multidisciplinary research questions. The second was surveying across community capital transects (social, cultural, built, economic, and public health), an approach grounded in "capital"-based resilience theory to promote the capacity for replication across a wide range of communities and hazards. Here, we describe our data collection strategy design and development process and  challenges in its implementation.

## Street View Data for Disaster Reconnaissance

The NHERI RAPID Facility, housed at the University of Washington, provides investigators with equipment, software, and support services to collect, process, and analyze perishable data from natural hazards events (Berman et al. 2020). The RAPID Facility is part of the National Science Foundation-supported Natural Hazards Engineering Research infrastructure (NHERI). It is a distributed, multi-user, national facility that provides the natural hazards engineering community with access to research infrastructure, combined with education and outreach, to support knowledge and innovation to prevent natural hazard events from becoming societal disasters. The RAPID Facility's science plan supports convergence research by emphasizing collecting multidisciplinary data across temporal, geospatial, and social scales.

Among its equipment, the RAPID Facility houses two vehicle-mounted, high-resolution iSTAR Pulsar+ mobile imaging systems to support post-disaster investigations. The iSTAR Pulsar+ can be used by the hazards and disaster research community to quickly and non-invasively collect high-resolution "street view" imagery data across a large cross-section of affected areas. Longitudinal (time-series) data collected post-event can provide insights on the community's recovery and the impact of response and recovery strategies.

Over the past decade, dozens of research investigations have shown that street view imaging "*can be mined for creative place-based metrics*" (Campanella 2017) pertinent to social science (Gebru et al. 2017), health science (Curtis et al. 2013), natural science (Swanger and Admassu 2018), and engineering (Curtis and Mills 2012). Street view imaging can be a low-cost technique to quickly acquire high-resolution, ground-based data across large areas. For example, Figure 1 shows a fused wide perspective composite image acquired in early April 2020 during testing of the RAPID Facility's new iSTAR system. Visible in this single, time-registered scene are (from left to right) an empty bus stop, three pedestrians in a crosswalk, one stopped car at a traffic light, a lighted red sign indicating an open business, and a sign indicating the business type.

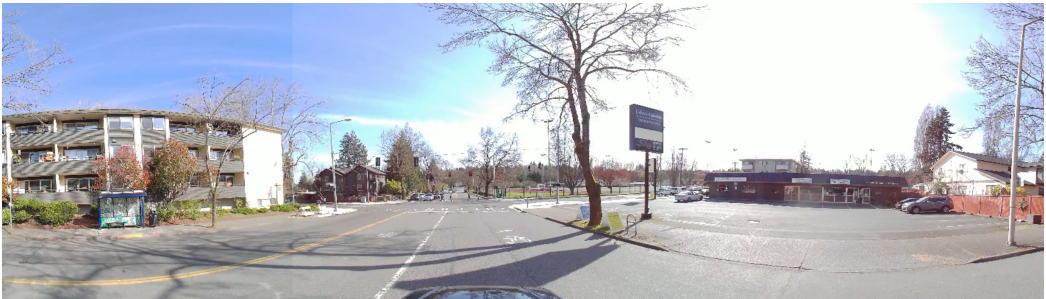

*Figure 1: Example of street view imagery acquired in Seattle in early April 2020 using the RAPID Facility iSTAR Pulsar+ system.*

The RAPID Facility's iSTAR Pulsar+ mobile imaging system has been used by several investigators to collect imagery data following multiple hazards (e.g., 2018 Camp Fire and 2020 Hurricane Laura). These data collection efforts have been mainly designed to support investigator-driven research without much regard to how other investigators use the data to answer research questions by other NHERI users. As such, the utility of the collected data to the broader hazards and disaster research community is limited, as is its potential to promote learning across hazard events.

## Initial Route Design

We employed two strategies (broad neighborhood/area canvassing and community capitals transects) to create an approximately 100-mile route through the city of Seattle, designed to take approximately 8 hours to complete (Figure 2).

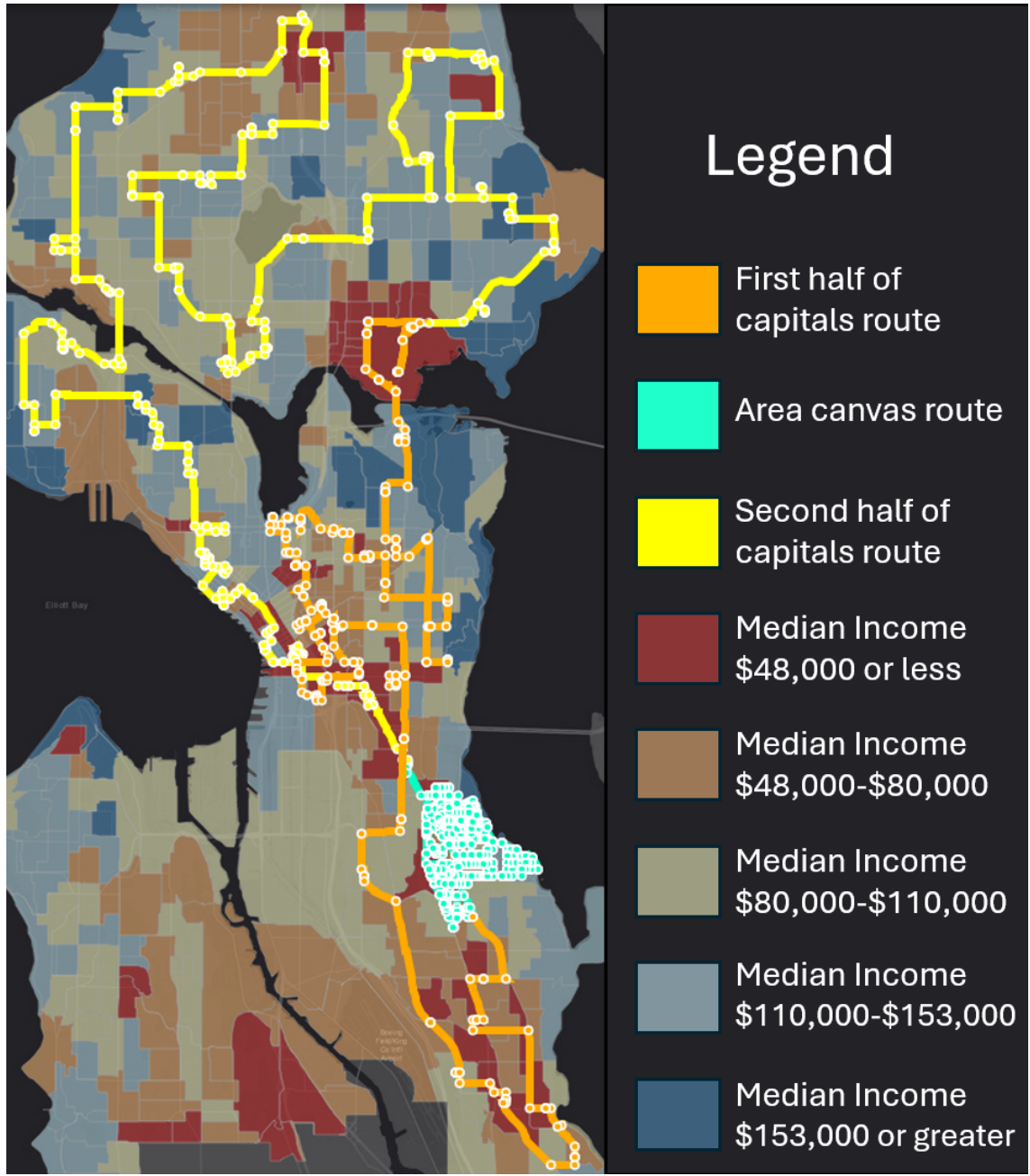

*Figure 2: Route utilized during the data collection campaign and Seattle median income 2014-2018 American Community Survey (ACS) 5-year estimates by census block group.*

Strategy 1: Broad neighborhood/area canvassing: This umbrella approach was designed to facilitate collection of data to answer multidisciplinary research questions, including about infrastructure and community wellbeing. This section of the route covers nearly all intersections in an area with a spread of income levels.

Strategy 2: Community capitals transects: We grounded our second approach in "capital"-based resilience theory to promote the capacity for replication across communities and hazards (Miles 2015). We selected five capitals (social, cultural, built, economic, and public health [a pandemic-specific adaptation of human capital]) relevant to the current disaster, and identified components (i.e., assets) specific to each of these capitals that were thought to be susceptible to pandemic impacts at the time of route design  (Table 1). Using publicly available data (e.g., licensed healthcare facility information, food establishment inspection data, Seattle tourism information), we identified locations with centroids within 200 feet of the road network associated with these components.  This 200-ft distance is a compromise allowing for buildings with large footprints to be included in the route, while not overestimating the visibility of smaller capitals from the car-mounted camera. We then created a transect that connects component locations in a way that is optimized for driving time.

**Table 1: Event-relevant Assets used for Route Design**

| Capital | Event-Relevant Assets | Example Data Sources |
|---|---|---|
| Social | Faith-based organizations, emergency shelters, libraries | Association of Religious Data Archives (Roger Finke et al. 1997), Seattle Libraries (City of Seattle 1995) |
| Cultural | Parks, museums, sporting and entertainment venues, community centers | King County Landmarks (King County GIS Center 2017) |
| Built | Construction, gas stations | Seattle Building Permits (City of Seattle 1995), King County Land Parcels (King County GIS Center 2017) |
| Economic | Grocery stores, farmers markets, restaurants | King County Land Parcels, King County Restaurant Inspections (King County GIS Center 2017) |
| Public Health | Hospitals, healthcare and medical facilities, pharmacies, opiate treatment centers, mental health providers, needle exchanges | Public Health Data (Washington State Department of Health 2016), King County Public Health Data (King County GIS Center 2017) |

Route design Process: We used an iterative approach to operationalize the two campaign design strategies into a single eight-hour route. First we needed to identify event-relevant components to generate capital transects.

We identified specific components of each capital that, based on our observations as residents and expertise as hazards and disaster researchers, had ongoing or expected impacts due to the pandemic or associated restrictive public health measures. After we identified capital transects of interest, we collected geolocation data for each capital in addition to 2014-2018 American Community Survey (ACS) 5-year estimates for median household income at the block group level.

       We loaded all the collected data into the ArcGIS Pro software. This software is capable of creating a route through the city that minimizes drive time according to our specifications using the Network Analyst toolbox and ArcGIS StreetMap Premium data. Initially, we had the software make a route that started and ended at the UW RAPID Facility and reached every identified capital in the city from our data sets. As expected, the generated route took multiple days to complete as it covered such a large portion of the city. To trim down the route to a reasonable length, we identified capital component types that covered a large geographic area of Seattle, experimenting with different combinations until we generated a route that reached all major areas of the city. The capital component types included in this route were hospitals, schools, museums, and opioid treatment programs. After confirming the resulting route visually met the need of reaching many different parts of the city, we identified other capital locations along the route that were not included in the generation. Additionally, we hand-edited parts of the route to maximize capitals coverage and minimize three-point turns. After multiple iterations,  a 5.5-hour route that reaches as many capitals as possible was achieved.

       The above methods achieved our goals for community capitals transects, but we still needed to add more to the route to meet our goals for neighborhood canvassing. To do this, we clustered the ACS median income data into 5 groups using Jenks natural breaks optimization. Then we visually identified an area which appeared to have a wide spread of income levels and road types. We synced the start and end points of the canvas route with waypoints already along the capitals route. We used all intersections in the selected canvas area in the ArcGIS Pro software to solve the route, resulting in a 2.5-hour route. To remain within the eight hour route limit, we hand-edited the route to reduce three-point-turns along the edges of the canvas area. After replacing the overlapping part of the capitals route with the canvas route, we finalized a 7.5-hour drive-time route that passes within 200 feet of over 2000 capital components in the city.

       <u>Ethical considerations</u>: Prior to data collection, we consulted with the University of Washington's Human Subjects Division to determine if our campaign would be subject to institutional review board (IRB) review. We

were advised that since people out in public cannot expect privacy, they are not considered human subjects. As such, IRB review would not be required. While this determination was based on interpretation of the statutes that govern human subjects research in the United States, we felt that the ethical considerations were more nuanced and that potential participant privacy concerns needed to be addressed. We decided to blur faces and license plates prior to public release of any data to respect the privacy of individuals who may be included in our images.

## Campaign Frequency and Route Revisions

With the initial route design in place, we made plans for regular data collection surveys utilizing the route. However, it became apparent that changes to the route were necessary after the first data collection effort. Additionally, issues beyond our control led to modifications to the campaign frequency and timing.

Route Revisions: After the initial survey, we discovered some problems with the route. While we had attempted to minimize three-point-turns in the initial route creation, there were still a few present in the route. These turns were required when the route included dead-end roads or driveways. Each turn cost the driver multiple minutes to complete and did not add much to the route in terms of capturing data at additional capital components. Thus we manually adjusted the route a final time to avoid all three-point turns, so that the driver could complete the route more quickly. This would become especially important as traffic increased when pandemic restrictions were lifted.

Campaign frequency strategy: We endeavored to collect data approximately once every 2 weeks for months 1-4 and once every 6 weeks for months 5 and beyond. In addition, we planned to collect data in response to major changes in public health recommendations and the implementation of community-based non-pharmaceutical interventions (NPIs). For example, we would plan to commence a survey in the days following the relaxation of stay-at-home or shelter-in-place orders, and then resume our scheduled missions at planned intervals. Surveys would be conducted on the same day of the week (Friday) and at the same start and stop time for consistency. We also had some plans to perform additional surveys at nighttime.

Campaign frequency reality: Several external factors affected our ability to adhere to the above campaign strategy. The first was the availability of the iSTAR Pulsar+ mobile imaging system. In addition to camera malfunctions requiring overseas shipment for repairs, the RAPID Facility's equipment lending plan dictates that immediate post-event field investigations are prioritized over long-term and longitudinal investigations. Weather events were another factor that affected data collection efforts. The iSTAR Pulsar+ mobile imaging system, like Google's own systems, cannot be used in the rain. This resulted in some data collection surveys being performed on days of the week other than Friday. Additionally, after initial testing, we determined that the nighttime images were too low quality to pursue the nighttime surveys as planned.

## Challenges and Discussion

In addition to the difficulties discussed in the preceding section, the implementation of our campaign introduced several challenges related to data publication, visibility of components used to develop capital-based transects, community engagement, changing traffic patterns, and more. Here, we describe our experiences, along with lessons learned and recommendations for future research and practice where applicable.

Data publication: We had intended to process, curate, archive, and publish these data within the NHERI DesignSafe-CI (Rathje et al. 2017) immediately following each survey. We also planned to post the processed data on an online web-mapping service. However, we encountered several challenges that delayed data publication due to the sheer quantity of data produced. The data generated from each survey exceeds one terabyte, requiring extended time to post-process and transfer data to offsite servers. Given pandemic restrictions limiting the amount of time staff spent in the office, our ability to upload data was delayed.

We had multiple options to consider for online web-mapping services. The two leading contenders were Mapillary and Google Street View. Google Street View required submitted images to be pre-anonymized, which given the quantity of data we were uploading, represented a challenge in validating in-house anonymization results. This made Mapillary the logical choice, as they would handle anonymization themselves. When submitting data to Mapillary however, we encountered decreased upload speeds, possibly due to throttling because of the size of the uploads.

In addition to storing the data on an online web-mapping service, we wanted to store the data on NHERI DesignSafe-CI with the goal of eventual publication on the site. While our data is stored on DesignSafe now, it is not yet published as it is not anonymized. Given the many users of DesignSafe, some of whom may want to upload and publish similarly large datasets that can overburden DesignSafe's data storage service, there are additional concerns about publishing a data set of this magnitude. Thus a larger policy discussion is needed to ensure consistent and ethical publication of the imagery data.

Visibility of capitals: Our preliminary review of the data collected indicated that several capital components used to generate the capital transect route were not visible, or were only partially visible, in the images. Moreover, we have observed that public data classifications of each of these capital components were not always fit for our purposes. For example, street medians under the care of the city's parks department were classified as "parks" in the public data. In result, these "parks" were given equal consideration as other "real" parks in our transect creation process. It is possible that our final route would be different if these misclassified capital components were removed from the data during the route design phase, although extensive data validation would be too time-consuming in most disaster reconnaissance missions.

Community engagement: While we worked with the University of Washington media relations team to broadly advertise our campaign, we regret that resource and time constraints and pandemic-related research restrictions prevented us from working directly with communities and local officials to design our campaign. We have since engaged in discussions with local government officials on potential applications of our data. Recognizing the importance of community engagement from study outset, we recommend investigators planning to collect similar imagery data in disaster-affected communities engage with local officials, community members, and leaders, for example by hosting listening sessions.

Changing traffic patterns: Our route was designed to take approximately eight hours to complete. Our protocol called for starting data collection at the same time each survey so that we would collect images at each component around the same time. At the beginning of our campaign, the stay-at-home order significantly limited the number of cars on the road. As restrictions have been lifted and more cars are on the road, traffic has extended our campaign time and hindered time-stamp consistency of imagery at each component. Additionally, there are variations in day-to-day traffic and weekday versus weekend traffic, which also cause variations in location timestamps.

Other considerations: During our campaign, community-based demonstrations in response to the deaths of several Black Americans at the hands of police occurred along our route. The vehicle operator was also advised by protesters that they would prefer if the vehicle avoided the area. We respected the wishes of these individuals and made the decision to modify our route while these protests were ongoing.

## Conclusion and Recommendations

While we encountered challenges, our approach was successful in generating longitudinal image data that can be used for a variety of research purposes in the hazards and disaster research community. Accordingly, we hope that these methods can serve as a blueprint for researchers to apply for future disaster events. Based on our experience, we recommend a few steps for future researchers to improve on our methods

First, we suggest collection of baseline (pre-event) data when possible, for example in communities with high disaster risk. We do not have survey data from before the COVID-19 pandemic. Publicly accessible data (e.g., from Google Street View or Mapillary) are generally too inconsistent for meaningful comparison (e.g., images are taken using different mobile imaging devices at different temporal resolutions for different locations). As such, it is difficult to conduct statistical inference comparing pandemic conditions to pre-pandemic conditions. If possible, we would recommend that researchers develop their route and collect baseline data ahead of time in disaster-prone areas. In addition to having baseline data, this practice will also allow researchers to fix any issues with route designs ahead of time.

Additional recommended r improvements include ground-truthing data sources used in route design, confirming survey equipment availability, and putting in place a data publication plan ahead of time. Ground-truthing data sources will avoid the mislabeling issues we discovered and help design a route as intended. If possible, confirming survey equipment availability throughout the lifecycle of a longitudinal survey campaign can facilitate more data collection on consistent days or times. While weather may ultimately require rescheduling, having access to the equipment on desired survey days will enhance the likelihood of being able to complete the campaign as planned. Finally, given the large amount of data created by this process, having an anonymization and publication plan in place ahead of time will allow for timely publication of the data generated by surveys. This allows for fast access to the data for your research group and others who may also use the data.

The COVID-19 Seattle Street View Campaign represents an unprecedented, high-resolution, ground-based record of the urban region during and after a major global crisis. It covers both community capitals transects and a broad neighborhood canvas to provide a holistic view of the city and how it responded to COVID-19. Rather than be the only data set of its kind, our hope is that providing our methods here will allow this data set to become the first of many. Generating more data of this type can promote learning across hazard events and locations, and create data products that can be utilized by researchers from a variety of disciplines.

## Acknowledgments

We gratefully acknowledge the U.S. National Science Foundation for funding support (NSF grant CMMI-2031119); Kevin Ramsey at BERK Consulting, Inc. for GIS consulting support for route design; Jaqueline Zdebski, Andrew Lyda, and Michael Grilliot at the NHERI RAPID Facility for operations support (e.g., survey implementation, data processing, sharing of on-ground insights and feedback for future research); Ciara Gormley for NSF REU-sponsored research participation and support; Mapillary for streaming the anonymized street view survey data (findable by searching 'Seattle, Washington' on the Mapillary platform and filtering by the username 'uwrapid'); and DesignSafe and the Texas Advanced Computing Center (TACC) at The University of Texas at Austin for providing the cyberinfrastructure including extensive data storage.  Any opinions, findings, and conclusions or recommendations expressed in this material are those of the author(s) and do not necessarily reflect the views of the National Science Foundation.